\documentclass[twocolumn,prl]{revtex4}
\usepackage{amssymb,bbm,bm}
\usepackage{graphicx}
\usepackage{bm}
\usepackage{amsmath}
\usepackage[usenames]{color}
\usepackage{epsfig}
\usepackage{hyperref}
\usepackage{subfigure}
\usepackage[sans]{dsfont}
\hypersetup{colorlinks=true, citecolor=blue,
linkcolor=blue,urlcolor=blue }

\begin{document}

\title{Localization and delocalization in one-dimensional systems with translation-invariant hopping}
\author{Reza Sepehrinia}\email{sepehrinia@ut.ac.ir}
\affiliation{Department of Physics, University of Tehran, Tehran 14395-547, Iran}
\affiliation{School of Physics, Institute for Research in Fundamental Sciences, IPM, Tehran 19395-5531, Iran}

\begin{abstract}
We present a theory of Anderson localization on a one-dimensional lattice with translation-invariant hopping. We find by analytical calculation, the localization length for arbitrary finite-range hopping in the single propagating channel regime. Then by examining the convergence of the localization length, in the limit of infinite hopping range, we revisit the problem of localization criteria in this model and investigate the conditions under which it can be violated. Our results reveal possibilities of having delocalized states by tuning the long-range hopping.
\end{abstract}
\maketitle

According to well-known theories of Anderson localization \cite{anderson1958absence,mott1961theory,abrahams1979scaling},
single-particle wave functions
are exponentially localized in low dimensional ($d=1,2$) disordered systems. Several
mechanisms have been identified which provide counterexamples to this belief \cite{evers2008anderson}. One route to delocalization is long-range hopping which especially in systems with one-dimensional (1D) geometry is feasible for systematic analytical treatment. It has been a useful model to investigate various properties of the Anderson transition and new aspects of it are still being discovered \cite{nosov2019correlation}.

The effect of long-range hopping on localization was first considered by Anderson \cite{anderson1958absence} and subsequently, in the problem of phonon localization, by Levitov \cite{levitov1989absence}. The following picture has emerged: for a hopping amplitude decaying as $1/r^{\alpha}$ with distance $r$, all states are extended if $\alpha<d$, whereas for $\alpha>d$ the states are localized. This was well confirmed by the power-law random banded matrix model \cite{mirlin1996transition} which describes a 1D system with random long-range hopping. This model undergoes an Anderson transition with multifractal eigenstates at $\alpha=1$. It turns out however that the above picture is not universal and does not hold for the models with correlated hopping \cite{bogomolny2003spectral,ossipov2013anderson,celardo2016shielding,nosov2019correlation}. The latter includes the models with correlated random hopping and non-random hopping with the on-site disorder. It is found that correlated hopping tends to localize the states even when $\alpha<d$. Regarding the extended states in these models, there have been few reports although the corresponding energies form a set of null measure. \cite{rodriguez2000quantum,rodriguez2003anderson,de2005localization,deng2018duality}.

In this paper, we report an analytical study of the localization properties of a class of correlated models characterized with translation-invariant hopping and diagonal disorder. Our approach is to start from the arbitrary finite-range hopping, for which we are able to obtain the localization length, and then take the limit of the infinite hopping range. This leads us to reconsider the criterion of localization in systems with long-range hopping and discuss the conditions under which it can be violated. Our results reveal possibilities of having delocalized states in these systems based on the asymptotic behavior of hopping.

\textit{Model}.
The model under consideration is one-dimensional tight-binding chain, represented by the eigenproblem
\begin{equation}\label{model}
   \sum_{n=1}^r t_n(\Psi_{i+n}+\Psi_{i-n})+\epsilon U_i\Psi_i = E\Psi_i,
\end{equation}
with hopping range $r$ and weak random potential $\epsilon U_i$, where
$\langle U_i\rangle=0$ and $\langle
U_iU_j\rangle=\sigma^2\delta_{ij}$. Angular brackets denote the ensemble average. In the absence of random potential, the solutions of (\ref{model}) are plane waves with
energy
\begin{equation}\label{dispersion}
    E(k)=2\sum_{n=1}^r t_n\cos nk,
\end{equation}
where unit lattice spacing is assumed and the wave vector $k$ belongs to the Brillouin zone $k \in [-\pi,\pi]$.

\textit{Perturbation theory}.
The solution of (\ref{model}), in the presence of the weak random potential, can be treated perturbatively \cite{derrida1984lyapounov} by rewriting it in terms of variables $R_i=\frac{\Psi_{i+1}}{\Psi_i}$,
\begin{equation}\label{model-R}
    \sum_{n=1}^r t_n\left(\prod_{m=0}^{n-1} R_{i+m}+\prod_{m=1}^{n}\frac{1}{ R_{i-m}}\right)=E-\epsilon U_i.
\end{equation}
For an unperturbed plane wave, $R_i$ is constant and for a perturbed solution it is assumed to be weakly fluctuating around that constant value, which can be expressed as $R_i=A \exp(B_i \epsilon + C_i \epsilon^2 + \cdots)$. This assumption is valid if the unperturbed solution is a single plane wave. Otherwise scattering to other states will produce superposition of waves with different wavelengths and thus position-dependent $R_i$ \cite{sepehrinia2010irrational,xie2012anderson}. Therefore we will be considering the single-channel part of the energy band. Since the dispersion relation (\ref{dispersion}) is an even function of $k$, for a given allowed energy, there are at least, two solutions $\pm k$ i.e. one channel of propagation. In order to eliminate one of the two wave vectors $\pm k$ an infinitesimal imaginary part can be added to the energy and finally be made to approach zero. From Eq. (\ref{model-R}), up to second order in $\epsilon$, we have
\begin{subequations}\label{aBC}
\begin{eqnarray}
  && \sum_{n=1}^r t_n (A^n+A^{-n}) = E, \label{eq-a}\\
  && \sum_{n=1}^r t_n \left( A^n \sum_{m=0}^{n-1} B_{i+m} - A^{-n} \sum_{m=1}^{n} B_{i-m} \right) = -U_i, \label{eq-B} \\
  && \sum_{n=1}^r t_n \left\{ A^n \left[\sum_{m=0}^{n-1} C_{i+m} + \frac{1}{2} \left(\sum_{m=0}^{n-1} B_{i+m} \right)^2 \right] \right. \nonumber \\
  && \hspace{1cm} + A^{-n} \left. \left[ -\sum_{m=1}^{n} C_{i-m} + \frac{1}{2} \left( \sum_{m=1}^{n} B_{i-m} \right)^2 \right] \right\} = 0. \nonumber\\ \label{eq-C}
\end{eqnarray}
\end{subequations}
Equation (\ref{eq-B}) can be written in closed form,
\begin{equation}\label{B2}
\sum_{m=1}^r (\tau_m B_{i+m-1} - \tau^*_m B_{i-m}) = -U_i,
\end{equation}
where $\tau_m=\sum_{n=m}^r t_n A^n$ and $\tau^*_m=\sum_{n=m}^r t_n A^{-n}$. As $E$ approaches the eigenenrgies of the pure system, $A$ becomes pure phase and thus $A^*=A^{-1}$ and therefore $\tau^*_m$ will be complex conjugate of $\tau_m$. The Lyapunov exponent (LE) and its weak disorder expansion is given by
\begin{eqnarray}\label{LE}
    \gamma(E)&=&\lim_{N\rightarrow \infty}\frac{1}{N}\sum_{i=1}^N \log R_i=\langle \log R\rangle\\
    &=&\log A +\epsilon \langle B\rangle+\epsilon^2\langle C\rangle+\cdots.
\end{eqnarray}
In order to calculate the averages, we take the average of equations (\ref{eq-B}) and
(\ref{eq-C}), from which we obtain
\begin{eqnarray}
  \langle B\rangle &=& 0, \\
  \langle C\rangle &=& - \frac{1}{2}\frac{\sum_{n=1}^r \varrho_n t_n (A^n+A^{-n})  }
  {\sum_{n=1}^r n t_n (A^n-A^{-n})}, \nonumber \\ \label{mean}
\end{eqnarray}
where $\varrho_n=\rho(0) + 2\sum_{l=1}^{n} (n-l) \rho(l)$ and $\rho(\tau)$ is the autocovariance function $\langle
B_{n+\tau}B_n\rangle$. The covariances should be obtained using Eq. (\ref{eq-B}).
By multiplying $B_{i+j}$ in Eq. (\ref{B2}) for $j=-r, -r+1, \cdots , r-1$ and using the symmetry $\rho(l)=\rho(-l)$ and $\langle B_{i+j}U_j \rangle=0$ for $j<r-1$ (because of statistical independence), we obtain the following set of $2r$ linear equations
\begin{eqnarray}\label{set}
\sum_{n=1}^r [\tau_n \rho(|n-1-j|) - \tau^*_n \rho(|n+j|)] = -\frac{\sigma^2}{\tau_r}\delta_{j,r-1},\nonumber \\
j=-r, -r+1, \cdots , r-1.
\end{eqnarray}
This is a linear inhomogeneous system to obtain $2r$ unknowns $\rho(0), \rho(1),\cdots,\rho(2r-1)$. The solution of this system for arbitrary $r$ does not seem to be simple. Without explicitly solving the equations, we were able to construct the numerator in Eq. (\ref{mean}) by linear combination of them. The final result is the closed expression for the average
\begin{equation}\label{}
\langle C \rangle = \frac{-\sigma^2}{2 \left[\sum_{n=1}^r (\tau_n - \tau^*_n)\right]^2},
\end{equation}
and the localization length (inverse LE) follows from it
\begin{eqnarray}\label{loc-length}
   \xi &=& -\frac{2}{\sigma^2\epsilon^2}\left[
   \sum_{n=1}^r n t_n (A^n-A^{-n})\right]^2,\\
   &=& \frac{2v^2}{\sigma^2\epsilon^2};\hspace{0.3cm} v=-2\sum_{n=1}^r n t_n \sin nk, \label{groupv}
\end{eqnarray}
where we have used $A=e^{ik}$. We can see from Eq. (\ref{dispersion}) that $v=\partial E/\partial k$ is the group velocity. At the band edges where the group velocity vanishes the Lyapunov exponent diverges which implies the failure of the analytic expansion in disorder strength \cite{derrida1984lyapounov}. The result (\ref{groupv}) implies that in the single propagating channel regime the states will be localized if
\begin{equation}\label{crit}
\sum_{n=1}^r n t_n \sin nk < \infty.
\end{equation}
As we can see, this condition always holds for the finite range $r$.

We now consider
infinite-range hopping and see if (\ref{crit}) holds in the limit $r\rightarrow \infty$ or not. A necessary condition for convergence of the series is $nt_n\rightarrow 0$ as $n\rightarrow \infty$. The first conclusion which can be drawn from this is that in order to have a localized state the hopping integrals should necessarily decay faster than $n^{-1}$. This result is indeed the Levitov's criterion of localization for $d=1$. However, the above condition is not a sufficient condition of convergence. Below we will see the cases for which the above condition holds, but the series does not converge. Before that, we state a more strict condition of convergence. It is known from the theory of trigonometric series \cite{bari1964treatise} that cosine and sine series, in (\ref{dispersion}) and (\ref{groupv}), with monotonically decreasing coefficients, are convergent except, perhaps, at $k=0$. Therefore, if the hopping decays faster than $n^{-1}$ but \textit{monotonically} then the series converges and the states will be localized. We now apply the general result Eq. (\ref{groupv}) to specific examples that have been studied before by other means.

\begin{figure}[t]
\epsfxsize8truecm \epsffile{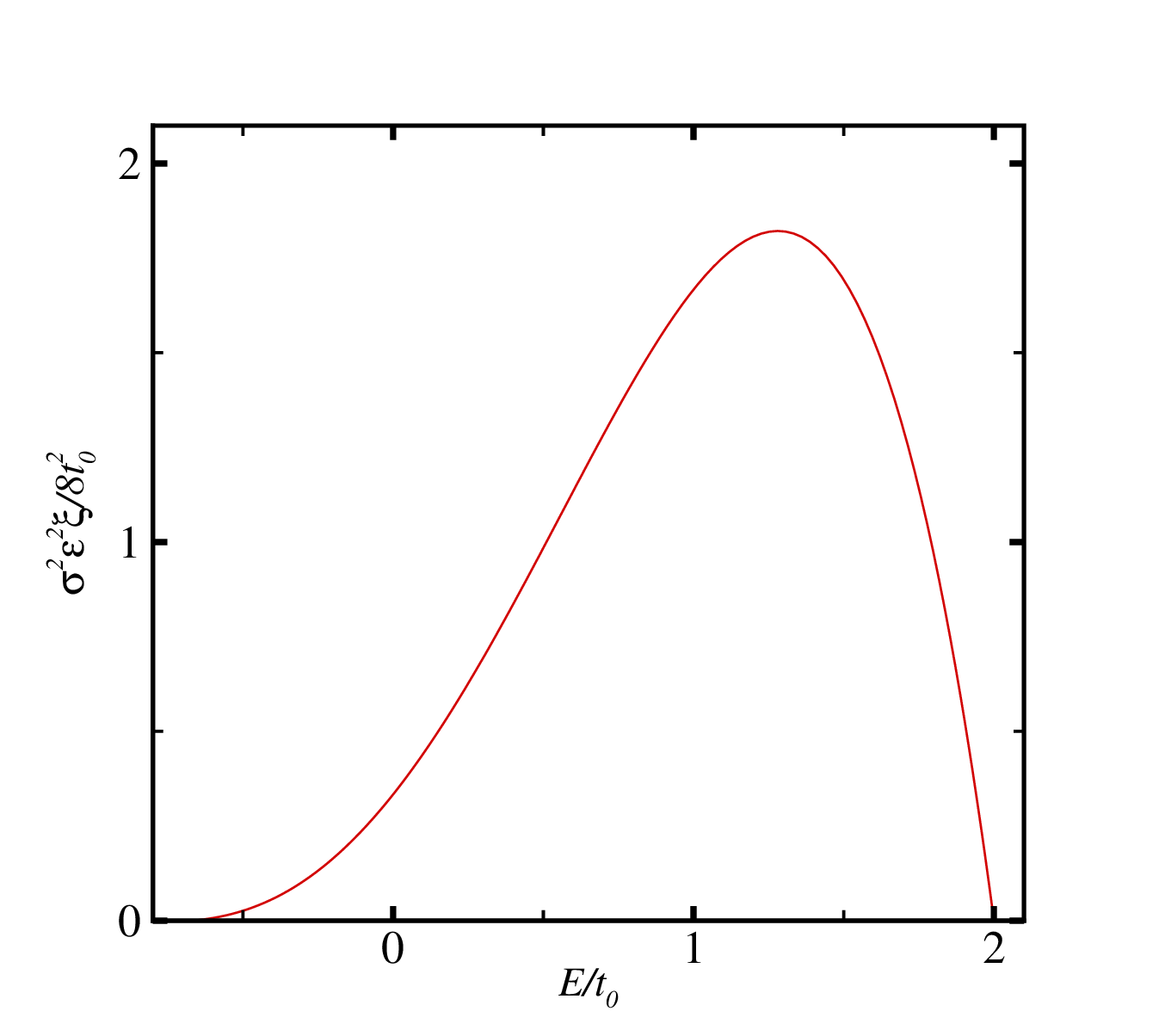} \caption{(Color online) Localization length for exponential hopping model $t_n=t_0 s^n$ with $s=1/2$.}\label{exp}
\end{figure}

\textit{Exponential hopping}. First we consider $t_n=t_0 s^{n}$ with $|s|<1$, we have
\begin{eqnarray}\label{exp-dis}
   E(k) &=& \frac{ 2t_0 s(\cos k-s)}{1-2s\cos k +s^2}, \\
   v(k) &=& \frac{2t_0 s(s^2-1)\sin k}{(1-2s\cos k+s^2)^2}. \label{exp-groupv}
\end{eqnarray}
From (\ref{exp-dis}) we can see that, for a given energy, there is only one pair of wavevectors $\pm k$ i.e. there is only one propagating channel and from (\ref{exp-groupv}) we can see that for all $k$ the localization length is finite and thus the corresponding states are localized. This model is studied in Ref. \cite{biddle2011localization} by numerical calculation of the inverse participation ratio (IPR). In agreement with their conclusions, our results show weakly localized states at higher energies (see Fig. \ref{exp}). As the range of hopping becomes shorter ($s\rightarrow 0$) the results tend to that of nearest-neighbor hopping Anderson model. Random band matrix model with exponential hopping, which is closely related to (\ref{model}), exhibits similar localization properties \cite{fyodorov1991scaling}.

\textit{Power-law hopping}. A more interesting case is $t_n=t_0 n^{-\alpha}$
which has been studied in several works \cite{rodriguez2000quantum,rodriguez2003anderson,de2005localization,celardo2016shielding,biddle2011localization} and shown to exhibit anomalous localization properties. In this model as well the dispersion relation allows a single channel of propagation, so our results are applicable,
\begin{eqnarray}\label{power}
   E(k) &=& t_0 [\text{Li}_{\alpha}(e^{ik})+\text{Li}_{\alpha}(e^{-ik})]; \hspace{0.2cm} \alpha>0,\\
   v(k) &=& it_0 [\text{Li}_{\alpha-1}(e^{ik})-\text{Li}_{\alpha-1}(e^{-ik})]; \hspace{0.2cm}\alpha>1,
\end{eqnarray}
where $\text{Li}_{\alpha}(z)=\sum_{n=1}^{\infty}z^n n^{-\alpha}$. We distinguish three different cases:

(\textit{i}) $0<\alpha \leq 1$. The series $E(k)$ converges for all $k$ except $k=0$ (band edge) where it diverges to infinity, $E(k\rightarrow0)\rightarrow +\infty$, so the energy spectrum of pure chain is not bounded from above. However, since $nt_n$ is not decreasing, the series $v(k)$ does not converge and in fact it is oscillating as $r\rightarrow \infty$ so the localization length does not have a well defined limit. This signals the failure of the assumption of exponential localization.

\begin{figure}[t]
\epsfxsize8truecm \epsffile{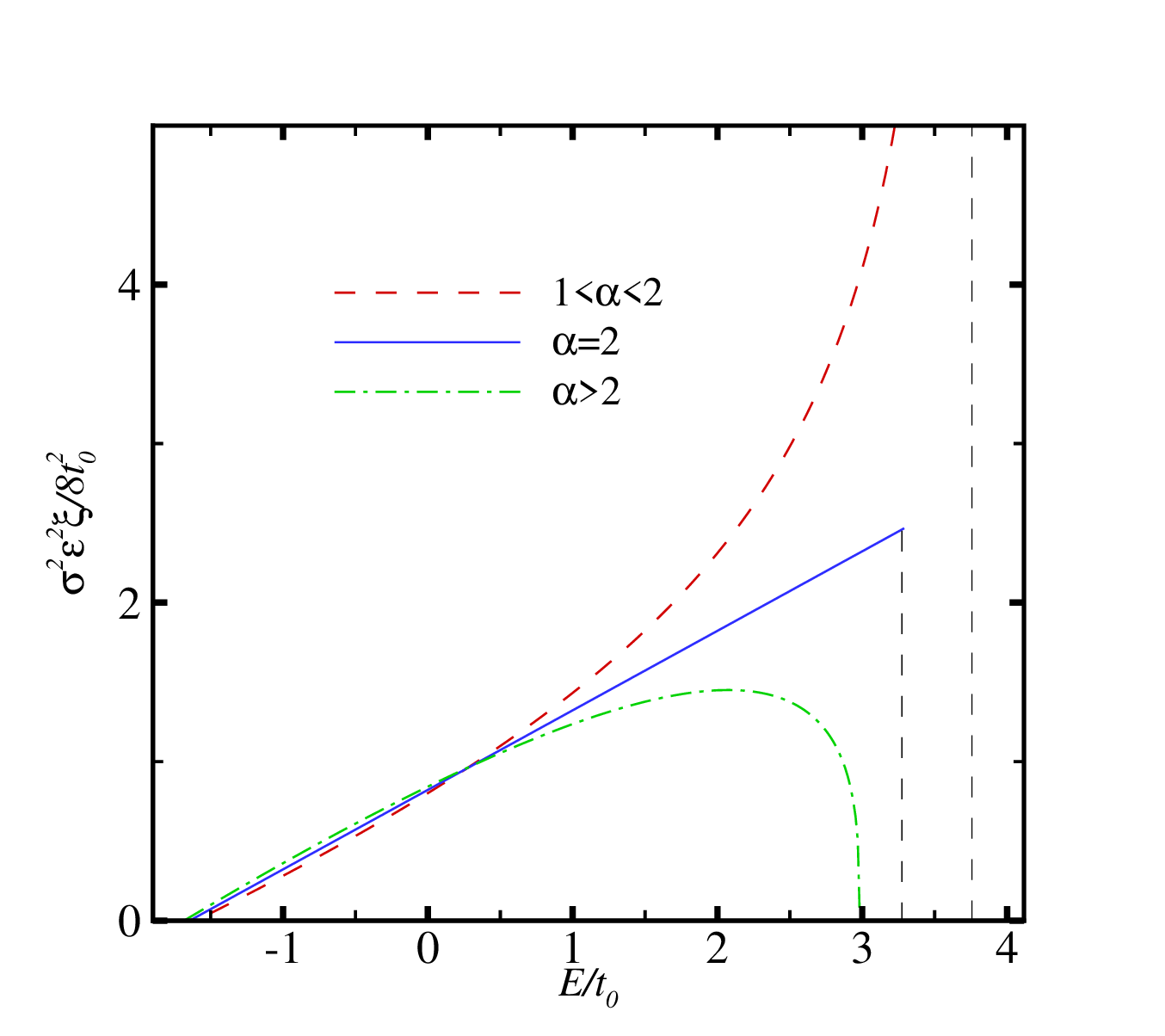} \caption{(Color online) Localization length versus energy for power-law hopping $t_n=t_0 n^{-\alpha}$. Vertical dashed lines show the upper band edge in each case.}\label{Fig2}
\end{figure}

(\textit{ii}) $1<\alpha <2$. The series $E(k)$ converges everywhere, including $k=0$, therefore the energy spectrum of the pure chain is bounded i.e. the bandwidth is finite. The series $v(k)$ also converges for all $k$, thus all states have a finite localization length for this range of $\alpha$. Although the localization length is finite, it increases unboundedly close to the upper band edge indicating delocalized states (see Fig. \ref{Fig2}). Delocalization of uppermost states has been predicted in Refs. \cite{rodriguez2000quantum,rodriguez2003anderson,biddle2011localization} and their transition to localized states at strong disorder is studied in Ref. \cite{de2005localization}. However, we do not see a qualitative change of behavior at $\alpha=3/2$, as is predicted in Ref. \cite{rodriguez2003anderson} and the power-law localization of states in this power-law hopping model (see Ref. \cite{nosov2019correlation}).

(\textit{iii}) $\alpha \geq 2$. Both the series $E(k)$  and $v(k)$ are convergent and bounded for all $k$ (see Fig. \ref{Fig2}). This confirms the numerical results of Ref. \cite{biddle2011localization} where it is found that there is a minimum IPR for this range of $\alpha$.

\textit{Delocalized states}. We now look for the sequences of hopping integrals for which the localization length diverges i.e. the condition (\ref{crit}) is not satisfied. The divergence of such a series is an old problem in the theory of trigonometric series \cite{zhizhiashvili2002pointwise,bari1964treatise} and is also related to the theory of functions with divergent Fourier series.

\begin{figure}[t]
\epsfxsize8truecm \epsffile{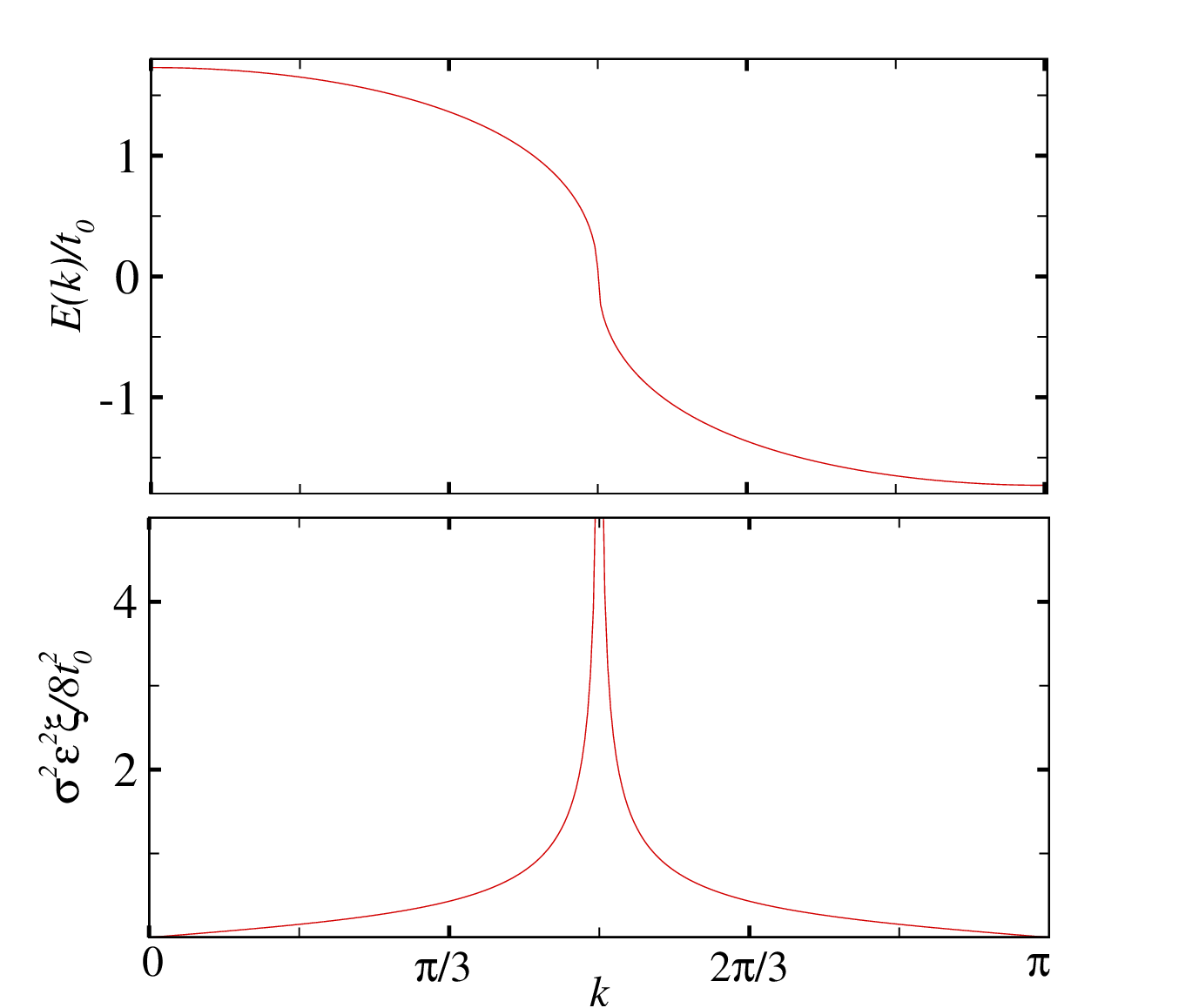} \caption{(Color online) Dispersion relation (top) and Localization length (bottom) for modulated power-law hopping $t_n=t_0 n^{-\alpha}\sin nk_0$ with $1<\alpha<2$ and $k_0=\pi/2$.}\label{k0}
\end{figure}

We have already seen that for power-law hopping with $1<\alpha < 2$ the localization length increases unboundedly near the upper band edge. By modulating the hopping with a sine wave $t_n=t_0 n^{-\alpha} \sin nk_0$ this singular point can be shifted into the energy band. This allows us to have an extended state at a given energy $E(k_0)$ and, in particular, the band edge can be avoided because the perturbation theory fails at this point. Such an oscillating hopping can be induced by RKKY interaction. The divergence of the localization length manifests itself in the dispersion curve as an infinite slope i.e. infinite group velocity (see Fig. \ref{k0}). This kind of singularity also occurs in the dispersion curve of Hartree-Fock excitations in interacting electron system. By superposition of multiple terms, $t_n=t_0 n^{-\alpha} \sum_i \sin nk_i$, we will have a set of extended states at given wavevectors. Particularly, this can be a dense set of energies at any given interval through the energy band.

In general, for $t_n=a_n \sin nk_0$ where $a_n$ is monotonically decreasing but $\sum_{n=1}^{\infty} a_n=\pm\infty$, there will be an extended state at $k_0$. We note that the single channel condition on the dispersion relation also needs to be satisfied. As an example $a_{n\geq2}=t_0(n\ln n)^{-1}$ can be considered. The nearest-neighbor hopping should be large enough such that the dispersion relation satisfies the single channel condition. Note that hopping decays faster than $n^{-1}$ but due to nonmonotonicity, condition (\ref{crit}) is not satisfied at $k=k_0$.

Finally, we would like to point out the possibility that extended states form a continuous band rather than a set of isolated energies. It is known that with certain (decreasing) coefficients the trigonometric series in (\ref{crit}) diverges almost everywhere. A suitable example for our discussion is $t_n = a_n \sin nq_n$ with certain conditions imposed on the sequences $a_n$ and $q_n$ \cite{zhizhiashvili2002pointwise,o1959divergent}. An explicit choice is $a_{n\geq 2}=t_0(n \ln n)^{-1}$ and $q_{n\geq2}=\ln \ln n$. Again the nearest-neighbor hopping should be such that the single channel condition is satisfied. We also note that for this choice $E(k)$ converges. The other interesting case would be the divergence of the series in a subinterval which results in a band of extended states separated by a mobility edge from the localized states; we leave this to future work.

\textit{Conclusions}. An analytical expression for the localization length in a one-dimensional tight-binding model with diagonal disorder and arbitrary-range hopping in the single channel regime is obtained. Finite-range hopping always leads to localized states but delocalized states emerge in the infinite-range limit. It turns out that for infinite-range hopping, $t_n \lesssim n^{-1}$ is a necessary but not sufficient condition for localization. We provide examples which satisfy this condition but violate the condition (\ref{crit}) and lead to delocalized states. The additional requirement of monotonic decay makes it a sufficient condition. Exponential and power-law hoppings were investigated in detail, and a qualitative comparison with previous studies was done. Our results reproduce several aspects of existing results although we arrive at different conclusions in some cases. Namely, for power-law hopping with $\alpha\leq1$ the localization length does not converge at $r\rightarrow \infty$, therefore the assumption of exponential localization seems to be invalid. Also, contrary to the predicted transition at $\alpha=3/2$, our results do not indicate a qualitative change at this point.

\section{aknowledgement}
We would like to acknowledge financial support from the research council of the University of Tehran for this research.

\bibliography{myref}
\bibliographystyle{apsrev}

\end{document}